# TITLE PAGE

# Building a Disciplinary, World-Wide Data Infrastructure

**Authors**

Françoise Genova, Christophe Arviset, Bridget M. Almas, Laura Bartolo, Daan Broeder, Emily Law, Brian McMahon

**Affiliations**

Françoise Genova, CDS (WDS Regular Member), Observatoire Astronomique de Strasbourg, UMR 7550 Université de Strasbourg/CNRS, France, ORCID 0000-0002-6318-5028

Christophe Arviset, Chair of the IVOA (WDS Network Member), ESA, European Space Astronomy Centre, Madrid, Spain, ORCID 0000-0002-2311-740X

Bridget Almas, Software Architect, Perseids Project, Tufts University, 5 the Green, Medford, MA 02155, USA

Laura Bartolo, Center for Hierarchical Materials Design, 2205 Tech Drive, Hogan 1160, Northwestern University, Evanston, IL 60201-2919  USA

 Daan Broeder, Meertens Institute Oudezijds Achterburgwal 185, 1012 DK Amsterdam

Emily Law, Deputy Program Manager, NASA/ Jet Propulsion Laboratory; President of Earth Science Information Partners (ESIP)

Brian McMahon, International Union of Crystallography, 5 Abbey Square, Chester CH1 2HU, UK

**Corresponding author**

Françoise Genova	francoise.genova@astro.unistra.fr

**AUTHOR INFORMATION**

Bridget Almas is the software architect and co-director of the Perseids Project at Tufts University. Bridget has worked in software development since 1994, in roles which have covered the full spectrum of the software development life cycle, focusing since 2007 in the fields of digital humanities and pedagogy.  Bridget served as an elected member of the Technical Advisory Board (TAB) of the Research Data Alliance (RDA), from 2013-2015, and currently is co-chair of the Research Data Collections Working Group and acts as liaison between the Alliance of Digital Humanities Organizations (ADHO) and RDA.

Christophe Arviset is the Head of the Data and Engineering Division at ESA's European Space Astronomy Centre, Madrid, Spain. Christophe has been involved in archive and data management for almost twenty years, heading the ESAC Science Data Centre (ESDC). The ESDC hosts the science archives of all ESA astronomy, planetary and solar heliospheric satellites. Christophe currently serves as the Chair of the International Virtual Observatory Alliance and is also ESA's representative at the International Planetary Data Alliance Steering Committee.

Laura Bartolo is a Senior Research Associate at Northwestern University and the Data/Database Efforts Coordinator at the NIST Center of Excellence for Hierarchical Materials Design (CHiMaD), a collaboration of Northwestern University, the University of Chicago, and Argonne National Laboratory.  Her background is in Information and Library Science and Laura has been involved in materials science and engineering data management and retrieval efforts over the past 15 years. She is the current Co-Chair of the RDA/CODATA Materials Data, Infrastructure, and Interoperability Interest Group and a member of the Materials Resource Registry Working Group.

Daan Broeder has a background in Electrical engineering and Computer Science and was for many years senior developer for archive and infrastructure solutions. He was the CTO and deputy head of the TLA archive at the MPI for Psycholinguistics and was involved in the coordination of many EU and national research infrastructure projects. He was responsible for creating infrastructure standards related to PIDs and metadata for language technology and resources in ISO TC37/SC4. Currently he is affiliated with the Meertens Institute of the Royal Dutch Society doing technical coordination within the Dutch CLARIAH project and is responsible for the community engagement within the EUDAT project.

Françoise Genova is an astronomer. She had been director of Strasbourg astronomical data centre CDS from 1995 to 2015, leading the data centre transition to the Internet era. She has been one of the founding parents of the IVOA, and serves as a member of the Board, representing France and Europe. She chairs the IVOA Data Curation and Preservation Interest Group. She coordinated several projects funded by the European Commission to develop the European Virtual Observatory. In parallel, she has been involved in the Research Data Alliance, its Technical Advisory Board and the RDA Europe projects from the beginning. She currently serves as co-chair of the RDA TAB.

Emily Law has over twenty years of experience in research, development and management of complex information systems. Emily currently serves as Deputy Program Manager and Development Manager at NASA/Jet Propulsion Laboratory to two separate directorates covering data systems in solar system research and earth science. She also leads operations for NASA's Planetary Data System, and oversees the development of data management infrastructures and portals in support of Earth and Planetary science. Since 2010, she has been an active member of the Earth Science Information Partners (ESIP) and is currently serving as the ESIP President.

Brian McMahon is Research and Development Officer of the International Union of Crystallography (IUCr). He works in the development of IUCr online journals and data publication services, and has been involved for 25 years in the Crystallographic Information Framework (CIF) data exchange standard programme.

**ABSTRACT**

Sharing scientific data with the objective of making it discoverable, accessible, reusable, and interoperable requires work and presents challenges being faced at the disciplinary level to define in particular how the data should be formatted and described. This paper represents the Proceedings of a session held at SciDataCon 2016 (Denver, 12-13 September 2016). It explores the way a range of disciplines, namely materials science, crystallography, astronomy, earth sciences, humanities and linguistics, get organized at the international level to address those challenges. The disciplinary culture with respect to data sharing, science drivers, organization, lessons learnt and the elements of the data infrastructure which are or could be shared with others are briefly described. Commonalities and differences are assessed. Common key elements for success are identified: data sharing should be science driven; defining the disciplinary part of the interdisciplinary standards is mandatory but challenging; sharing of applications should accompany data sharing. Incentives such as journal and funding agency requirements are also similar. For all, social aspects are more challenging than technological ones. Governance is more diverse, often specific to the discipline organization. Being problem-driven is also a key factor of success for building bridges to enable interdisciplinary research. Several international data organizations such as CODATA, RDA and WDS can facilitate the establishment of disciplinary interoperability frameworks. As a spin-off of the session, a RDA Disciplinary Interoperability Interest Group is proposed to bring together representatives across disciplines to better organize and drive the discussion for prioritizing, harmonizing and efficiently articulating disciplinary needs.

# MAIN TEXT

## 1. Introduction

A growing number of scientific communities are improving their organizational abilities to share data in their own and, increasingly, neighbouring fields. High-quality disciplinary building blocks are no less essential for cross- and inter-disciplinary research than they are in pursuing disciplinary research. Such building blocks contribute to the ideal of Findable, Accessible, Interoperable and Reusable (FAIR) Data (Wilkinson et al. 2016). It is important to share lessons learnt by disciplines with extensive experience and more established practice with those that are starting the process and still have to develop their own strategy. A key point within each discipline is finding out how and where to organize any coherent discussion of disciplinary interoperability. In some subjects, there are national, regional or international mechanisms for doing this. However, science is a global endeavour, and the full power of data sharing is to make data 'open to the world'. Thus interoperability discussions are not necessarily best managed from a local or national level perspective. An area in which there are few coordinated activities is the mapping out and understanding of commonalities and differences in the approach of different disciplines.

The excellent UK Data Archive guide to Managing and Sharing Data (Van den Eynden et al. 2011; see also Corti et al. 2014) provides an effective summary of best-practice guidelines based on wide-ranging input from many disciplines. However, it has a national focus, and is perhaps more exhortative than analytical. We do not know of any detailed survey results that give a clear indication of the uptake of such best-practice advice across many disciplines and worldwide.

Thus, we considered that an excellent opportunity to examine this question - of how a number of different disciplines in practice build their international disciplinary data sharing frameworks - was the SciDataCon 2016 conference, organized by CODATA (the ICSU Committee on Data) and WDS (the World Data System, another organisation affiliated to ICSU). This meeting was part of International Data Week (12-17 September 2016, Denver, CO, USA), which also included the 8[th] Plenary meeting of the Research Data Alliance. The present paper arose from a session designed to gather data specialists from different disciplines to exchange perspectives on their data sharing practices. This session, called 'Building a disciplinary, world-wide data infrastructure', took the form of a number of brief presentations and a panel discussion around a specific set of questions. This approach does not represent a large-scale survey; but the panel representatives are all experienced in the strategic development of data management activities within their disciplines, and so could bring a critical synoptic overview of their communities' practices to the table.

A range of disciplines (or sub-disciplines) participated, representing different cultures: humanities, linguistics, astronomy, earth sciences, materials sciences, and crystallography. In the case of humanities and material sciences, it was decided to invite a subdiscipline, linguistics and crystallography respectively, to also present their views, to illustrate the diversity of approaches within broad disciplinary fields. Comparison of different disciplinary ways and means of enabling and developing an interoperability infrastructure for each community was encouraged with the following set of questions:

1. What is your disciplinary culture with respect to data sharing?
2. What are the science drivers for building your discipline interoperability infrastructure?
3. How is your discipline organized to build its interoperability infrastructure?
4. What are the lessons learnt from building your discipline interoperability infrastructure?
5. Which part of your discipline interoperability infrastructure could be made generic for other disciplines?

The elements provided for each discipline, mostly following this analysis scheme are summarized in Section 2. The presentations were followed by a discussion including questions from the audience, which brought additional topics such as interdisciplinary data sharing or the usage of 'generic' e-infrastructures. Commonalities and differences in the disciplinary approaches are summarized in Section 3, and conclusions in Section 4.

## 2. Disciplinary Examples

### a. Humanities

It is difficult, if not impossible, to provide a comprehensive answer to the proposed questions for the entirety of the humanities, as practices within the sub-disciplines of the humanities can be as diverse as those of the individual sciences.  Nonetheless, some general observations can be made.  The culture of data sharing in the humanities follows a variety of (often ad hoc) approaches: via linked data URIs, the Application Programming Interfaces (APIs) of Content Management Systems and custom websites, GitHub repositories, Google drive, Dropbox, etc.  More formalized approaches, such as institutional repositories, shared infrastructure services, Dataverses, etc. are used when they are available and accessible, but there is not a single common practice across the entire domain of the humanities and availability of such services is not universal (Padilla 2016).

Common drivers for data sharing are even more elusive, particularly where the cost/benefit ratio is unclear.  Directives from funders for data management plans, and awards given to projects that put a priority on cross-project collaboration and best practices around data publication do play a role; altruistic motives and the desire to provide pivotal resources for a given community can also inspire action.  Within the humanities, individual sub-disciplines organize themselves around shared, domain-specific infrastructures that solve specific problems and provide common services targeted at the domain's needs. We name a few specific examples of these focused infrastructures. Perseids is an infrastructure for analysis and annotation of ancient texts. Papyri.info is a platform for editing, publication and interoperability across different papyrological resources. NINES (Networked Infrastructure for Nineteenth-Century Electronic Scholarship) is an organization and collection of tools and services for research into nineteenth century scholarship. Symogih is a collaborative platform providing a general model for historical research. Links to the web sites of these projects and the other ones cited in this paper are provided at the beginning of the Reference list.

In Europe and Australia, national (and cross-national in Europe) general purpose infrastructures for humanities data are also available (Lossau 2012). This is largely not the case in the United States, however. The fact that often the largest obstacles to data sharing are not technical, but instead have to do with questions of governance, funding and sustainability, is one important lesson to be taken from humanities infrastructure and data sharing practices.

University libraries and Information Technology departments may also need new models to support increasing demand for infrastructure and shareable data (Dinsman 2016). The humanities have valuable contributions to make towards general cross-domain infrastructure in the form of its use cases and requirements. The humanities can also contribute software, APIs and services from existing infrastructures. As more common cross-domain infrastructure components are developed, developers in the humanities contribute to the sustainability of such components by sharing the burden for their development and maintenance.

### b. Linguistics

The linguistics community has a long experience with standardization efforts for achieving interoperability of its research data and research data descriptions e.g. the ISO 639 suite of standards for language codes. But it was the establishment in Europe of the CLARIN research infrastructure, in the context of the European Strategic Forum Research Infrastructure (ESFRI) process which allowed work on developing a comprehensive research infrastructure to take place. The ESFRI process resulted in the implementation of research infrastructures in many disciplines. One outcome was the CLARIN European Research Infrastructure Consortium (ERIC) as a coordinating body that is supported by different European national CLARIN projects.

Although CLARIN was born within the linguistics community (Varadi et al. 2008), it was clear from the start that CLARIN should not limit itself to targeting only linguistics research but that the main research objects and products ('language data': newspaper corpora, speech recordings, lexica etc. and 'language technology': e.g. translation, named entity recognition) are also very useful for the broad Humanities such as, for instance, history and media-studies which use language data and technology.

CLARIN was set up as a distributed interoperable infrastructure especially facilitating the easy creation, and sharing of resources. It relies on a large network of (currently) a few dozen CLARIN centres of varying capabilities that store and offer access to resources and services. Although CLARIN is a European infrastructure, it is in principle not limited to Europe as the US CLARIN centre at Carnegie Mellon University demonstrates. The CLARIN centres are subject to a process of certification with respect to compliance with the technical infrastructure requirements: the use of Persistent Identifiers (PID), publication of interoperable metadata for the hosted resources based on Component Metadata (CMDI) and providing access to resources to users via the CLARIN Authentication and Authorisation Infrastructure (AAI) federation. Next to that CLARIN is concerned with the sustainability of infrastructure requiring the crossover of data and services between centres when needed.

The CLARIN centre experts participate in the different national and European committees and task forces that work on different aspects of the necessary interoperability services and agreements. For instance there is a standardization committee working on recommendations for data formats, a legal committee to work on licensing matters and a CMDI task force that works specifically on the metadata standard. The CLARIN ERIC ideally wants to be only a lightweight coordination layer, managing the task forces and committees, pushing for broad community support for CLARIN, and, most importantly, coordinating the contributions from the national CLARIN projects to the European CLARIN infrastructure. Essential infrastructure services, such as for instance the CLARIN centre registry, are operated either by special CLARIN A centres or under the responsibility of the CLARIN ERIC itself.

Many parts of the technical infrastructure are meant to facilitate easy sharing of resources and overcome the different national and sub-discipline idiosyncrasies that existed before. CLARIN supports open access and open data. Nevertheless there will remain justified limitations for making all resources open, as there can be legal and ethical reasons to keep access to resources limited to some groups or individuals.

Next to building and maintaining technical infrastructure CLARIN also engages with researchers and students, to inform them about the CLARIN infrastructure and its use within research.

The key recommendations from CLARIN, that data centres need to assign PIDs and metadata to digital objects to make them discoverable, accessible, interoperable and re-usable, and the importance of data centre certification, are now widely recognized as bases for successful and sustainable scientific data sharing.

### c. Astronomy

In astronomy, the Virtual Observatory (VO) is the vision that astronomical data sets and other resources provided by projects and data centres worldwide should work as a seamless whole. The International Virtual Observatory Alliance (IVOA) is an organization that debates and agrees the technical interoperability standards that are needed to make the VO possible.

For a long time, astronomy has had an open data and data sharing policy. Usually after a 6-12 months proprietary period for observational data, astronomical data is made available by projects and data centres freely and openly to the scientific community through on-line web services (web based Graphical User Interfaces and scriptable APIs). Similarly, most of the software developed is released under open source licenses. In general, astronomy science has no borders and astronomers share a very collaborative culture, based on the fact that all large astronomy projects (ground- and space-based) are performed under large international consortia.

Since its inception, the IVOA has been science driven and this has been enhanced with the creation of the Committee for Science Priorities (CSP) which defines priority science cases to be supported by the VO framework, such as multi-wavelength science, combining data sets from multiple sources, time variability of objects, data discovery, data access, data analysis and visualization tools.

The IVOA was created in 2002, very soon after the beginning of the VO development. The first VO projects obtained funding from the European Commission, and in the USA and in the UK starting in 2001. It was understood very quickly that a global organisation should be created to enable collaboration between the funded VO initiatives and to manage the development of the disciplinary interoperability standards which are at the core of the VO framework. One point discussed at that time was the level of representation of VO projects as 'IVOA members' represented in the Executive Board. It was decided that members would be funded initiatives at the national and regional level, the latter allowing representation of the Euro-VO projects funded by the European Commission. The VO initiatives of intergovernmental organisations were also invited to join later on – the European Space Agency ESA is currently one of the IVOA members. One of the difficulties has been the different funding cycles of the major VO initiatives, but the VO actors agree of the necessity to maintain the IVOA and it has been successfully done over the years. The IVOA membership comprises 20 initiatives worldwide in 2016.

The IVOA does not have any formal funding, but benefits from the in-kind voluntary contribution of its members. With one representative from each project, the Executive Committee sets the vision of the IVOA, and defines overall priorities for the work. Its chair rotates every 1.5 years and is supported by a Deputy Chair. The Committee meets every 3-4 months either in person or by teleconference. Work within the IVOA is organized through Working Groups (WG), the goals of which are to define interoperability standards, and Interest Groups (IG), where experiences with VO technology are discussed and fed back to the WGs. Anybody in the network can participate in any of the WGs and IGs. The chair and vice chair of the WGs/IGs are nominated by the Executive Committee and form together the Technical Coordination Group.

All the IVOA network meets twice a year in an interoperability face-to-face meeting gathering around 80-120 people, organized and paid by one of the IVOA members. During the rest of the year, WGs and IGs work through collaboration tools (wiki, mailing lists, web pages, teleconferences, etc). The IVOA has defined a technical architecture of the VO framework which has been very stable since 2010 (Figure 1).

Building the VO astronomy interoperability infrastructure over the last 15 years has revealed some important lessons such as: (1) The infrastructure needs to be driven by Science, and supported by applications; (2) Collaboration amongst projects is difficult and time consuming but compulsory; and (3) Engagement and take-up from data centres and large projects is key.

Many of the IVOA standards are very specific to astronomy (in particular the ones related to data models, although for instance the IVOA work on its Provenance data model interests the relevant RDA Interest Group), but others are more generic and can potentially be re-used outside astronomy (*i.e.* VOTable – the Virtual Observatory Table format standard, TAP - Table Access Protocol, SAMP – Simple Application Messaging Protocol, VOSpace ('dropbox for VO'), Registry of VO resources, …). We observed indeed strong re-use from neighbouring disciplines such as Planetary Science, Astroparticles, Atomic and Molecular Data. The IVOA Architecture vision itself is generic and could inspire other disciplines.

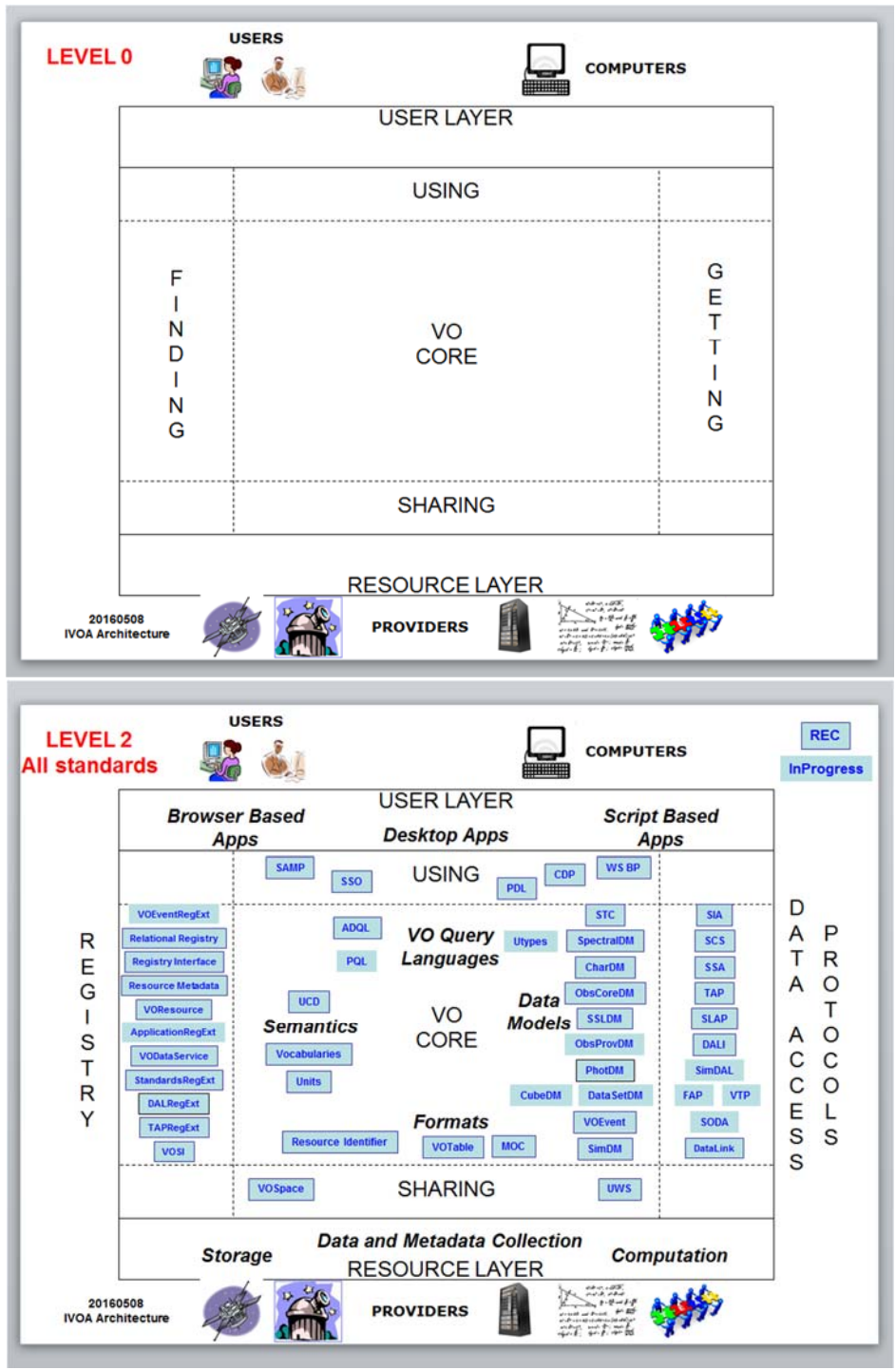

**Figure 1**: **The IVOA architecture** (Arviset et al. 2010). (top) Schema of the functions fulfilled in the interoperability layer. (bottom) Schema showing the topics dealt with by the IVOA and the recommendations (IVOA standards) relevant to the different topics (blue boxes).

### d. Earth sciences

The benefits of sharing Earth science data have been recognized during recent decades. Open data sharing is increasingly supported and its importance is widely acknowledged. During the past decade, the international scientific community has agreed on principles for sharing data on Earth observations. Effective science communication is another aspect of Earth Science's culture. It is gaining wide acceptance that alternative art forms can present research findings atypical of scientific journals and publications. For example, a 'Research as Art' event was held at the Earth Science Information Partners (ESIP) Summer 2016 Meeting and it was well attended and received. The event allowed the community to showcase their research methods and results in creative and artistic means.

Breakthroughs in understanding Earth phenomena are made when we are able to pursue broader and transformative interdisciplinary science. Grand challenges beyond a single discipline require integrated study for us to better understand the Earth System to predict outcomes, consequences and impacts. Building Earth data infrastructure in an innovative way that enables data, model and system interoperability is the key to integrate scientific data and knowledge stemming from various fields of research that are enabled by diverse Earth observations.

Most agencies manage their Earth science data through their silo disciplinary oriented data infrastructures (e.g. atmosphere, oceanography, solid Earth, etc.) Some data sets from the individual data centres around the world are discoverable through interoperable search services offered for example by the Global Change Master Directory (GCMD) and the Global Earth Observing System of Systems (GEOSS). There is considerable effort put forth in addressing interoperability worldwide through a variety of venues. Noticeably, ESIP is a community of data and information technology practitioners who come together to collaborate on coordinated interoperability efforts across Earth science communities. ESIP provides backend infrastructure that facilities collaboration on development of best practices and advancement of technology and interoperability.

Ensuring seamless interoperability is vital in advancing the use of Earth science data to increase Earth scientific discovery, and at the same time a tremendous challenge. Building a successful discipline interoperability infrastructure involves: (1) Clear understanding of science needs and use cases. (2) Well-developed roadmap with an architecture that takes into consideration interoperability aspects across data, models, services, applications and systems. (3) Making facilitated collaboration a priority. A voluntary inter-organization framework such as the one offered by ESIP maximizes collaboration. (4) Integrated solution and capability must address interoperability standards throughout Earth science data lifecycle and must produce quality data.

Lessons learned can be leveraged and are applicable to other disciplines in addition to established best practices, standards and technologies developed by specific discipline. Examples of what Earth science interoperability infrastructure can offer for broader disciplines include: (1) A high level common conceptual model comprises interoperable architectural elements including information architecture, process architecture and technology and service architectures. (2) ESIP collaborative framework, infrastructure and platform. (3) Technologies and standards such as semantics, web services, cloud computing, etc.

### e. *Materials Science*

To some, like the Strange Matter Exhibit, materials science is the 'study of stuff'. As presented by Jim Warren, NIST, at the SciDataCon 2016 session on 'Building a disciplinary, world-wide data infrastructure', materials science and engineering is an interdisciplinary field focused on the discovery and design of materials to improve the performance of products that impact our everyday lives, ranging from transistors, LEDs, titanium turbine blades, to sporting goods.

Materials research primarily examines the relationships between (1) the processing of a material, such as casting or polymerization; (2) the structure or morphology of a material, such as a fluid or a nanotube; and (3) the properties of a material such as reflectivity or hardness of a material. A **key challenge** facing materials science is to speed up the design of new materials and eliminate the slow process of unnecessary incremental repetition. The Materials Genome Initiative (MGI), a major US White House effort currently underway (Figure 2), is built upon creating and capturing knowledge of materials.

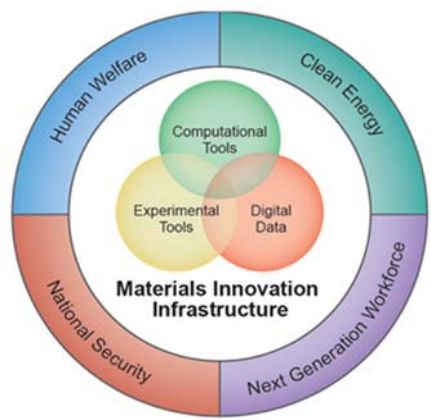

**Figure 2: The Materials Innovation Infrastructure (National Science and Technology Council 2014).**

With this foundation, materials innovation becomes a pipeline, derived from a deep understanding and control of material behaviour and properties. The ultimate aim of MGI is to accelerate advanced materials design in ways that significantly drive economic prosperity, address national and regional energy needs, and bolster national security (National Science and Security Council 2014). Improving this pipeline requires connecting efforts from a variety of independent but thematically congruent regional, national, and international materials design efforts to: align key stakeholders; connect diverse materials design and informatics expertise; simplify data access; and deploy modern scalable data services to support materials researchers by leveraging and synthesizing existing tools to form a cohesive, interoperable materials data infrastructure. The combined resources (human, data, computational) across the globe are such that we have a tremendous opportunity to accelerate progress towards MGI goals by leveraging common services and major materials data resources that are thus far disconnected and underused in order to improve integration of experimentation, computation, and theory for maximum benefits to the materials community (National Science and Security Council 2016).

At this time, a rapidly improving understanding of **data sharing** has taken hold in the materials research community. The connection between data sharing and increased research opportunities is gaining attention and support. However, as with other disciplines, data sharing in the materials community is hindered by a reward system that recognizes publication in high impact journals but minimizes the contributions of data and code sharing. While the materials software development community sees value in open code development, it must co-exist with the successful business model of commercial codes. Yet in all areas of the materials community, early career scientists show an openness and adoption of new approaches for data sharing.

Three primary factors are driving the materials community towards **building an interoperable** world wide data infrastructure: (1) The requirement in materials science to integrate experimentation, computation, and theory across length and time scales propels the need for better multiscale models (Minerals, Metals & Materials Society (TMS) 2015); (2) The need for accessibility of high-value data from beam lines (neutrons and photons) and supercomputers brings new research possibilities to materials scientists; and (3) The evident opportunities for long-tail big data are promising.

At this stage of infrastructure building in the materials community, several lessons have become apparent. While groups are supportive of a shared infrastructure, the changes needed for establishment are difficult to achieve. The need for better tools for building and adopting the infrastructure is widely supported. Finally standards are 'risky' and hard to reach consensus. NIST, in its efforts to support the Materials Genome Initiative, is leading standards development efforts, moving carefully with the materials community to construct, adopt, and implement needed standards to drive discovery and innovation.

### f. Crystallography

Crystallography is the study of molecular and crystalline structure and properties. Its most useful data output is the precise description of molecular structure, which is central to much of mineralogy, chemistry, physics, mathematics, biology and materials science. This may help to explain its traditionally open culture to data sharing. The infrastructure of crystallographic data is the curated structural databases (many of which are several decades old and built by academic scientists) and journals, especially those published by the International Union of Crystallography (IUCr), which apply high standards of technical review to data sets as well as to the literature. There is currently also a growing trend to link to the raw experimental data that is stored at the originating experimental facility or in institutional repositories.

Crystallographers have a strong desire to repeat and improve previously determined structures, to ensure the highest quality of published results, and to data-mine comprehensive archives of structural models. To facilitate these objectives, there is an overwhelming need for easy information flow and cross-linking between journals and databases, as well as among the various software packages used for data processing and analysis.

The databases describe a variety of types of structure, and have diverse architectures and indeed business models. However, there is a particular mechanism that allows data exchange between databases, software applications and journals. That mechanism is a data exchange protocol (CIF, the Crystallographic Information Framework) developed under the aegis of the IUCr (Hall, Allen & Brown 1991), again by practising academic scientists. It provides a complete description in machine-readable form of the concepts involved in a crystal structure determination. This includes precise data and metadata terms characterising the experiment, subsequent data processing, the structure solution and refinement, and the geometry and chemical characterization of the structural model. It forms, in effect, a machine ontology of the subject.

CIF was developed as an **extensible** ontology, allowing progressive interoperability between the components of the infrastructure to be built up gradually (Figure 3). Modular extensions to new types of structure or experiment have been added during the course of the past 25 years. The key to this approach is to be able to partition the task into discrete efforts that can be tackled by a group of domain experts, working within the semantic framework already established. The result is that there are now components of the CIF ontology that cover every element of the data workflow, from capture of raw experimental data, through data reduction, processing and analysis, model building to publication and database deposition.

Despite the clear advantages of a standard designed to facilitate interoperability, there was some inertia within the community to implement it in practice. An important – perhaps crucial – factor in overcoming this was the early decision of IUCr journals to mandate CIF as the submission vehicle for small-molecule structures, as well as the provision of basic tools to allow authors to obey this mandate. Practitioners within any discipline may need some such measure of compulsion to embrace any new standard.

Although crystallography has developed its standard ontology framework using a particular file format, the all-important definitions of data items and their attributes are contained in extensible format-agnostic dictionaries. Hence the approach is general, and could be adopted by any discipline (Hall & McMahon 2016). It has already been applied in some related areas of structural science, including nuclear magnetic resonance spectroscopy, 3-dimensional electron microscopy, chemical structure, materials properties and protein homology modelling.

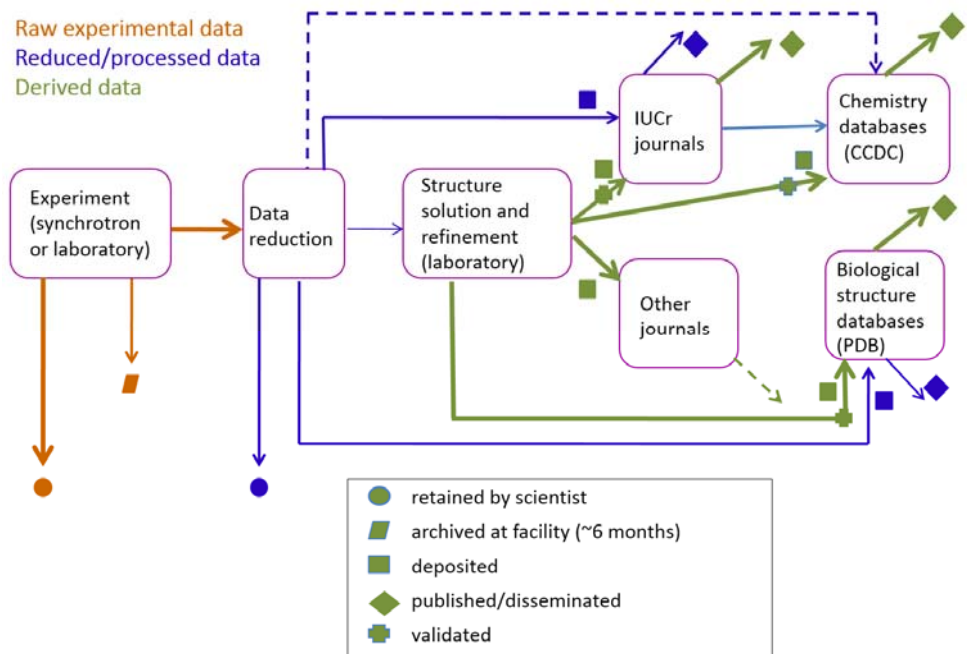

Figure 3. Schematic view of the implementation of the CIF data exchange standard (Hall & McMahon 2016). It illustrates how this implementation allows ready interoperability throughout the entire crystal structure determination workflow, from experimental apparatus through data reduction and analysis, model building, publication and database deposition.

3.  Commonalities and Differences

Several important **commonalities** appear among the different disciplines. Because of the variety of disciplines examined in this study, they are also key aspects of the development of a successful interdisciplinary data infrastructure.

(1) The most important one is certainly that **scientific data sharing endeavour must be science-driven**. This is recognized by all, and it is very important to keep it in mind because it means that technology should obviously be exploited, but to serve the scientific aims and not as a driver. Similarly, responding to top-down injunctions to share data or to use specific methods or support infrastructures is not enough to ensure that the data will be useful and usable.

(2) Defining the **discipline-specific part of interoperability standards**, in particular semantics and formats, is mandatory but can be difficult, since this requires at some stage international agreement and can involve different projects and organizations which may have their own aims. Such 'standardisation' discussions are also a social process where fostering broad support often requires some coordination by trusted community members.

(3) The question is **not only to share data, but also applications**, software, tools. It is essential, for community uptake of data sharing, that data producers are enabled to share their data as well as users enabled to use the shared data. Adoption by large projects, for the disciplines which implement such projects, and data centres is a key element of success.

(4) The well-known key question of **incentives** is also present. It appears that journal requirements can play a key role, directly linked to the importance of publications in researchers' careers. The collective power to solve common problems, to develop common best practices and standards and reusable software that stakeholders can bring back to their own organization clearly is a beneficial factor. The requirement by more and more funding agencies that projects provide a data management plan is also a major incentive, the effect of which should be more and more visible in the future.
(5) More generally, it is widely recognized that **social aspects are more challenging than technical ones**.

**Governance is more diverse**. The contributions of the different disciplines confirm that the ways disciplines organize themselves to establish their data interoperability framework can be different. Each discipline also has to manage the diversity of its sub-disciplines, which can have their own cultures and organizations.

Astronomy, a discipline well organized around large projects, in general built through international collaboration, used its practice of international collaboration to organize the development of its data interoperability framework.

The existence of strong international organizations in the field influences the earth sciences which use inter-organization frameworks to collaborate nationally and internationally to advance interoperability.

More fragmented disciplines, humanities and material sciences, also have to deal with huge data diversity. In these disciplines, sub-disciplines managed to define their interoperability framework: linguistics in Europe seized the opportunity to apply for membership in the ESFRI Roadmap; crystallography, which deals with highly structured data, is one of the pioneers of scientific data sharing and built its framework on a controlled vocabulary and shared data representations. Material sciences use discussion forums made available by organizations such as CODATA and the RDA. Grassroots approaches relying on informal agreements on common data models and use of shared service APIs are common in the humanities, but more formal commons-based models also spring up around specific areas of interest. Pelagios Commons is one such initiative, providing online resources and a community forum for open data methods for working with historical places.

**Many technical and technological aspects appear to be shareable**, including architecture concepts. The implementation of a registry of resources is essential to enable data discovery. It can rely on generic concepts with disciplinary extensions, such as the IVOA registry, based on OAI-PMH, which has been customized by nearby disciplines, and will be used as a model by the materials science community. The IUCr maintains a machine-readable registry of CIF ontologies and intends to extend this to include other compilations of metadata describing specific types of experiment.

There are technological and practical incentives, including gains in efficiency, to use **generic e-infrastructures**, for instance for storage, or **generic building blocks**, in particular for identity federation. Interoperability with existing disciplinary infrastructure can also be implemented. For instance, the IVOA registry is included in EUDAT B2FIND discovery service.

4. Conclusions

Disciplines which are not strongly organized at the international level may encounter more difficulties in establishing their interoperability framework. International organizations such as GEO, CODATA and the RDA can be used to fill this gap. International discussion forums can be established as CODATA Task Groups and/or RDA Interest and Working Groups. In addition, the WDS 'community of excellence' is a host for individual data providers as regular members, but also for disciplinary networks once they are constituted, and it organizes interaction between its members. The agriculture research community for instance set up an RDA 'Agriculture Data Interest Group (IG)' for discussion of the domain interoperability questions. The IG first initiated a Work Group (WG) to define elements of Wheat Data Interoperability, in the framework of the International Wheat Initiative. The WG produced a set of RDA recommendations on wheat data interoperability guidelines, ontologies and user cases. The next steps are a WG on Rice Data Interoperability built on the successful model set up by the wheat one, and a more general Agrisemantics WG. These Groups have a strong international participation and are linked to international organizations, such as the FAO. The RDA is a neutral forum for discussion. Participation in the RDA allows contacts with a wide range of experts of the very diverse building blocks of scientific data sharing, and with other disciplines, which is essential to work towards cross- and inter-disciplinary research and to assess the usage of generic elements of the data infrastructure for disciplinary needs.

Interdisciplinary research is essential in particular to work on the Grand Challenges. Environmental, earth or health sciences are for instance excellent use cases. They can span across a wide variety of disciplines, including social sciences and humanities. They should rely on high-quality 'disciplinary pillars', built with participation of the relevant communities. Lessons learnt from building disciplinary data interoperability frameworks show that it is essential that the building of bridges between the disciplinary pillars is problem-driven, not just an answer to top-down injunctions. The Cluster projects set up by the European Commission to establish synergies between large infrastructures such as the ESFRI ones also often have a strong component linked to data sharing and integration across 'data silos', and can include infrastructures from different disciplines, such as astronomy and astroparticle physics, two nearby disciplines with different data sharing cultures, in the Astronomy ESFRI and Research Infrastructure Cluster ASTERICS, or environment in ENVRI/ENVRIplus.

The RDA hosts many disciplinary Interest and Working Groups. The rich outcome of the SciDataCon 2016 session which gave rise to this paper suggests that it could be very useful to encourage these Groups to propose a 'Disciplinary Interoperability Framework Interest

Group', to host discussions among disciplines and produce statements of lessons learnt and best practices, on organisational and sociological aspects, in particular governance, as well as on more technological topics. This Interest Group could also organise discussions with the Groups which deal with common technological and sociological aspects of scientific data sharing, in particular around the disciplinary requirements and the relevance of the 'generic' outputs to the needs. It would also help the disciplines which already have an organisation to discuss their interoperability framework and do not see the need for a specific RDA Interest group on their own topic to engage fully in the RDA. The RDA is a unique Forum by the diversity of its members' profiles and of the disciplines represented, and likely the only place where these discussions can be organised with active participation of a variety of disciplines.

The suggestion to propose such an Interest Group as a spin-off of the SciDataCon session aroused immediate interest among RDA Group Chairs. Several of them, together with an IVOA member to represent disciplines which do not have a specific RDA Group, have now organised a so-called 'Bird-of-a-Feather' session at the RDA Barcelona Plenary in April 2017 to initiate the constitution of the Interest Group.

We conclude with a specific pointer towards one possible, more technology oriented spin-off activity. The results of this exchange of information and experience between discipline representatives are perhaps not very surprising. Relatively small communities focused on well-defined scientific areas with homogeneous observational or experimental practices already have well developed data exchange frameworks. In the case of astronomy, the standard image format FITS (Wells et al. 1981) has been continuously updated since its inception to become the widely adopted astronomy standard, the basic enabler of interoperability for astronomical data. Likewise, the Crystallographic Information Framework has diversified beyond its initial application to single-crystal diffractometry (Hall & McMahon 2016). In the digital classics community, Pleaides, a community-built gazetteer and graph of ancient places providing stable identifiers has served as an important pivot point to enable data sharing across projects (Elliott et al. 2014). However, the success of these technical approaches demonstrates the benefit of creating well-defined mechanisms that have specific utility, and subsequently encouraging other parts of the discipline community to extend and adopt such a solid working foundation to suit their own needs.

This is clearly going to be more difficult to achieve, the wider the area of interest encompassed by a discipline. Nevertheless, the "bottom-up' approach has the benefit of establishing core metadata that will be used across that discipline as its own data-handling skills mature. One of the greatest challenges to interoperability across disciplines is the 'gluing together' of separately-developed metadata catalogues. There exist very high-level 'top-down' metadata schemas such as Dublin Core (International Organization for Standardization 2009) which provide a useful perspective on global common elements, but disciplines develop their own extensions to the Dublin Core, as well illustrated by the diversity of the metadata schemes of OAI-PMH resource registries (see for instance the IVOA Registry Resource Metadata, Hanisch et al. 2007). Mapping of more granular metadata across the disciplinary divide will require concordances of terms and utilities for interconverting between file formats in which machine-readable metadata are conveyed. There are encouraging efforts to develop tools to facilitate this (see the disciplinary metadata catalogues of the Digital Curation Centre at http://www.dcc.ac.uk/resources/standards and the Research Data Alliance at http://rd-alliance.github.io/metadata-directory/). However, strategies for developing the necessary machine-processable concordances depend both on an awareness within the different disciplines of the need to devote resources to such activities, and an understanding of the real benefits that can follow from doing so. We believe that an interchange of experiences such as reported in this paper, and the on-going discussions in the RDA and its relevant outputs, can help to raise the necessary awareness and motivation.

**AKNOWLEDGEMENTS:**

James A. Warren's contribution in the discussion of Materials aspects during the panel is warmly acknowledged. John Westbrook, Ian Bruno, John R. Helliwell, Amy Sarjeant, Suzanna Ward and David Schade were co-authors of the initial contributions. Franciska de Jong, David Schade and Imma Subirats were co-conveners of the session.

**COMPETING INTERESTS:**

The authors declare that they have no competing interests.

**REFERENCES:**

ASTERICS. Available at https://www.asterics2020.eu/ [Last accessed 29 January 2017].

CLARIN. Avalable at https://www.clarin.eu/ [Last accessed 29 January 2017].

CODATA. Available at http://www.codata.org/ [Last accessed 29 January 2017].

ENVRIplus. Available at http://www.envriplus.eu/ [Last accessed 29 January 2017].

ESIP. Available at http://esipfed.org [Last accessed 29 January 2017].

ESFRI. Available at http://www.esfri.eu/ [Last accessed 29 January 2017].

EUDAT B2FIND. Available at https://www.eudat.eu/services/b2find [Last accessed 29 January 2017].

International Data Week 2016. Available at http://www.internationaldataweek.org/ [Last accessed 29 January 2017].

IVOA. Available at http://ivoa.net [Last accessed 29 January 2017].

GCMD. Available at http://gcmd.nasa.gov [Last accessed 29 January 2017].

GEOS. Available at http://www.earthobservations.org/geoss.php [Last accessed 29 January 2017].

NINES. Available at http://www.nines.org [Last accessed 29 January 2017].

NIST. Available at https://www.nist.gov/ [Last accessed 29 January 2017].

Papyri.info. Available at http://papyri.info [Last accessed 29 January 2017].

Pelagios Commons. Available at http://commons.pelagios.org/ [Last accessed 29 January 2017].

Perseids. Available at http://sites.tufts.edu/perseids [Last accessed 29 January 2017].

RDA. Available at http://rd-alliance.org [Last accessed 29 January 2017].

SciDataCon 2016 Session Building a Disciplinary, World-Wide Data Infrastructure. Available at http://www.scidatacon.org/2016/sessions/44/ [Last accessed 29 January 2017].

Strange Matter Exhibit. Available at www.strangematterexhibit.com [Last accessed 29 January 2017].

Symogih. Available at http://symogih.org/ [Last accessed 29 January 2017].

IUCr. Available at http://www.iucr.org/ [Last accessed 29 January 2017].

WDS. Available at http://www.icsu-wds.org/ [Last accessed 29 January 2017].

**Arviset, C, Gaudet, S and the IVOA Technical Coordination Group** 2010 IVOA Architecture Version 1.0. IVOA.

**Corti, L, Van den Eynden, V, Bishop, L, Woollard, M** 2014 Managing and Sharing Research Data: A Guide to Good Practice. London: Sage Publications Ltd.

**Dinsman, M** 2016 The Digital in the Humanities: An Interview with Laura Mandell - Los Angeles Review of Books. Available at https://lareviewofbooks.org/article/digital-humanities-interview-laura-mandell/ [Last accessed 29 January 2017].

**Elliott, T, Heath, S, Muccigrosso, J,** editors, 2014 *Current Practice in Linked Open Data for the Ancient World*. ISAW Papers 7 (2014). Available at http://dlib.nyu.edu/awdl/isaw/isaw-papers/7/ [Last accessed 29 January 2017].

**Hall, S R, Allen, F H and Brown, I D** 1991 The Crystallographic Information File (CIF): a new standard archive file for crystallography. *Acta Crystallogr*, A47: 655–685. DOI: http://dx.doi.org/10.1107/S010876739101067X

**Hall, S R and McMahon, B** 2016 The Implementation and Evolution of STAR/CIF Ontologies: Interoperability and Preservation of Structured Data. *Data Science Journal* 15: 3. DOI: http://doi.org/10.5334/dsj-2016-003

**Hanisch, R J, IVOA Resource Registry Working Group, NVO metadata Working Group** 2007 Resource Metadata for the Virtual Observatory Version 1.12. IVOA Recommendation 02 March 2007. ADS: 2007ivoa.spec.0302H

**International Organization for Standardization** 2009 ISO 15836:2009 - Information and documentation - The Dublin Core metadata element set. Geneva, Switzerland.

**Lossau, N.** 2012 An Overview of Research Infrastructures in Europe - and Recommendations to LIBER. *LIBER Quarterly*, 21(3-4):313–329. DOI: http://doi.org/10.18352/lq.8028

**Minerals, Metals & Materials Society (TMS)**, 2015 Modeling Across Scales: A Roadmapping Study for Connecting Materials Models and Simulations Across Length and Time Scales. Available at www.tms.org/multiscalestudy [Last accessed 29 January 2017].

**National Science and Technology Council (U.S.**). Subcommittee on the Materials Genome Initiative 2016 The Materials Genome Initiative: The First Five Years. Available at https://mgi.nist.gov/sites/default/files/uploads/mgi-accomplishments-at-5-years-august-2016.pdf [Last accessed 29 January 2017].

**National Science and Technology Council (U.S.).** Subcommittee on the Materials Genome Initiative 2014 Materials Genome Initiative Strategic Plan. **Padilla, T** 2016 Humanities Data in the Library: Integrity, Form, Access. *D-Lib Magazine,* 22.3/4. DOI: 10.1045/march2016-padilla

**Van den Eynden, V, Corti, L, Woollard, M, Bishop, L, Horton, L** 2011 Managing and Sharing Data: Best Practice for Researchers. UK Data Archive. Available at http://www.data-archive.ac.uk/media/2894/managingsharing.pdf [Last accessed 29 January 2017].

**Váradi, T, Krauwer, S, Wittenburg, P, Wynn, M and Koskenniemi, K** 2008 CLARIN: Common Language Resource and Technology. In Proceedings of the Sixth International Conference on Language Resources and Evaluation (LREC 2008), European Language Resources Association (ELRA), Marrakech, Morocco, May 28-30, 2008. Available at [http://www.lrec-conf.org/proceedings/lrec2008/pdf/317_paper.pdf](http://www.lrec-conf.org/proceedings/lrec2008/pdf/317_paper.pdf) [Last accessed 29 January 2017].

**Wells, D. C., Greisen, E. W., Harten, R. H.** 1981 FITS: A Flexible Image Transport System. *Astronomy and Astrophysics Supplement Series*. **44**: 363–370.

**Wilkinson, M D et al.** 2016 The FAIR Guiding Principles for Scientific Data Management and Stewardship. *Sci. Data*.**3**:160018. DOI: 10.1038/sdata.2016.18